# IMPACT OF CHANNEL PARTITIONING AND RELAY PLACEMENT ON RESOURCE ALLOCATION IN OFDMA CELLULAR NETWORKS

Sultan F. Meko

IU-ATC, Department of Electrical Engineering
Indian Institute of Technology Bombay, Mumbai 400 076, India
Email: sultanfeisso@ee.iitb.ac.in

*ABSTRACT*

Tremendous growth in the demand for wireless applications such as streaming audio/videos, Skype and video games require high data rate irrespective of user's location in the cellular network. However, the Quality of Service (QoS) of users degrades at the cell boundary. Relay enhanced multi-hop cellular network is one of the cost effective solution to improve the performance of cell edge users. Optimal deployment of Fixed Relay Nodes (FRNs) is essential to satisfy the QoS requirement of edge users. We propose new schemes for channel partitioning and FRN placement in cellular networks. Path-loss, Signal to Interference and Noise Ratio (SINR) experienced by users, and effects of shadowing have been considered. The analysis gives more emphasis on the cell-edge users (worst case scenario). The results show that these schemes achieve higher system performance in terms of spectral efficiency and also increase the user data rate at the cell edge.

*KEYWORDS*

*FRN deployment, frequency reuse, OFDMA, multi-hop, outage probability*

## 1. INTRODUCTION

Modern cellular networks provide various types of real-time and non real-time services. The amount of available resources (i.e., time-slot and subcarriers) and the Quality of Service (QoS) requirements determine the capacity of the cell while the transmit power and propagation conditions determine the cell size. Increased capacity, coverage and throughput are the key requirements of future cellular networks. To achieve these, one of the solutions is to increase the number of Base Stations (BS) with each covering a small area. But, increasing the number of BSs requires high deployment cost. Hence, a cost effective solution is needed to cover the required area while providing desired Signal to Interference plus Noise Ratio (SINR) to the users so as to meet the demand of the future cellular networks. To achieve the high data rate wireless services, Orthogonal Frequency Division Multiple Access (OFDMA) is one of the most promising modulation and multiple access techniques for next generation wireless communication networks. In OFDMA, users are dynamically allocated subcarriers and time-slots so that it is possible to minimize co-channel interference from neighboring cell by using different sub-carriers. Therefore, OFDMA based multi-hop system offers efficient reuse of the scare radio spectrum.

We consider an OFDMA-based cellular system in which users arrive and depart dynamically. Each arriving user demands rate $\bar{r}$. If the required rate can be provided, only then a user is accepted, otherwise it is blocked. Depending on the SINR experienced by an arriving user, the BS computes the subcarriers that are needed to be allocated to the user so as to provide the required rate. If the required sub-channels (i.e., a group of subcarriers) are available, then the user is admitted. Note that the SINR decreases as the distance between the BS and the user increases. Thus, the users at the cell boundary can cause blocking probability to be high. Since the number of admitted users is directly proportional to the revenue of the service provider, it is imperative to design solutions that allow accommodating a large number of users. This motivates us to propose a Fixed Relay based cellular network architecture that is well suited to improve the SINR at the cell boundary, and thus can possibly increase the number of admitted users. Relaying is not only efficient in eliminating coverage holes throughout the coverage region, but more importantly; it can also extend the high data rate coverage range of a single BS. Therefore bandwidth and cost effective high data rate coverage may be possible by augmenting the conventional cellular networks with the relaying capability.

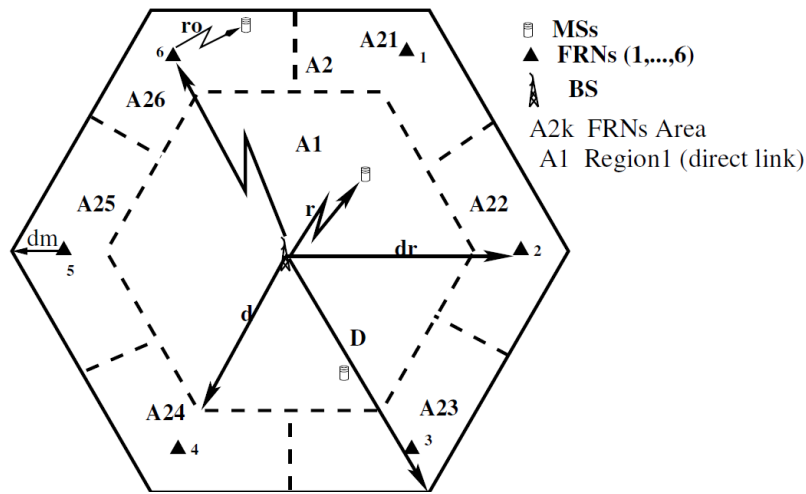

Fig.1. Layout of FRN enhanced Cellular system

We consider a cellular system with six fixed relay nodes (FRNs) that are placed symmetrically around the BS as shown in Fig.1. Mobile stations (MSs) in outer regions $A_{21}$ to $A_{26}$ can use relaying to establish a better path than the direct link to BS. The key design issues in such systems are the following: (1) How sub-channels should be assigned for (a) direct MS to BS links, (b) FRN to BS links, and (c) MS to FRN links. Algorithm for this is referred to as the *channel partitioning scheme.* (2) How sub-channels can be used across various cells. Algorithm for this is referred to as *channel reuse scheme.*

Effective channel partitioning maximizes utilization of every channel in the system, and thus obtains high spectrum efficiency in cellular systems [3]. For a cellular system enhanced with FRNs, the main idea of channel partitioning is to optimally assign the resources to the MS-BS, FRN-BS, and MS-FRN links. Such intra-cell spectrum partitioning along with the channel reuse scheme not only grant the data rate demanded by each user, but also manages the inter-cell interference by controlling the distance between any pair of co-channel links.

The concept of channel partitioning for FRN based cellular system has been discussed in [4], [5], [2], [8], [7], [13]. In [4], full frequency reuse scheme was proposed. The authors divided the cell in to seven parts and allocated six sets of subcarriers to FRN link and the remaining to BS. The authors aimed at exploiting the multiuser diversity gain. In [5], frequency reuse scheme was proposed based on dividing the outer region in to six and also sectoring the inner region. In [2], a pre-configured relay channel selection algorithm is proposed to reuse the channels that are already used in other cells on the links between FRNs and MSs. This scheme may suffer from high co-channel interference on FRN-MS links. In [7], [13], frequency partitioning and reuse schemes for cellular WLAN systems with mobile relay nodes are proposed. Relays are mobile as the MSs themselves act as a relay for other MSs. Since the relay is mobile, the channel between relay and the BS can change, which will result in a large number of inter-relay handoffs. Furthermore, MSs acting as relays may not be cooperative because of the power consumption and the security issues. In [8], a coverage based frequency partitioning scheme is proposed. The scheme assumes that the relay nodes are placed at a distance equal to the two-third of the cell radius, and does not consider optimal relay placement. Some other proposals for frequency management include the use unlicensed spectrum [11], and the use of directional antennas [1].

In our paper [6], we proposed a channel partitioning and a channel reuse scheme for increasing system capacity to support some preceding standards like Global System for Mobile Communications (GSM). In this paper, we extend the channel partitioning and channel reuse scheme for OFDMA cellular networks which results in increased system capacity and spectral efficiency. We also consider the optimal relay placement based on different parameters. We show that with the appropriate relay placement, the system performance can be improved significantly. As a result, the number of users that can be accommodated in the system can be maximized while providing each user with its required rate.

The paper is organized as follows. In Section 2, we describe our system model. In Section 3, we propose our channel partitioning and relay placement scheme based on different parameters. In Section 4, we evaluate the performance of the proposed schemes using numerical computations and simulations. In Section 5, we conclude the paper.

## 2. SYSTEM MODEL AND DESCRIPTION

### 2.1. System Configuration

We consider a cellular system consisting of regular hexagonal cells each of edge length D. Each cell has a BS and six FRNs situated symmetrically around the BS at a distance $d_r$ from the center and $d_m$ from the cell edge as shown in Fig.1. Let the total bandwidth available for the uplink is W units. Let the user density in the cell be λ, i.e., in a region of area A, λA MSs are present. Let each MS demands unit rate from the system. We assume the shortest distance routing scheme [9]. For a given MS position, let the distance from BS be $d^*$ and nearest FRN be $d^*_m$. If $d^* \leq d$, then MS communicates with the BS directly; otherwise it communicates through the nearest relay using two hop route. Because of the specified routing scheme, a cell can be partitioned into seven regions as shown in Fig.1. We define the region covered by BS as the inner region ($A_1$) and the region covered by FRNs as outer region ($A_2$). Outer region is further divided into $A_{2k}$, for k = 1, . . ., 6. All the MSs in region $A_1$ communicate directly to the BS, and the MSs in $k^{th}$ $A_2$ region communicate to BS through relay $k^{th}$ FRN.

### 2.2 Channel Partitioning and Reuse Scheme

The channel partitioning scheme, partitions the uplink bandwidth into thirteen orthogonal segments, viz. $W_1$, $W_{2,k}$ and $W_{3,k}$ for k = 1, . . . , 6. The band $W_1$ is used by the MSs in region $A_1$, the band $W_{2,k}$ is used by $k^{th}$ FRN to communicate with the BS, and the band $W_{3,k}$ is used by the MSs in $k^{th}$ $A_2$ region to communicate with $k^{th}$ FRN. Let $W_2 = \sum_{k=1}^{6} W_{2,k}$ and $W_3 = \sum_{k=1}^{6} W_{3,k}$. Because of the channel partitioning, there is no intra-cell interference, and the system performance is mainly determined by inter-cell co-channel interference.

For channel reuse scheme, we assume that the frequency reuse distance is 1, i.e., each cell uses the complete bandwidth *W* for the uplink communication [10]. This may cause significant co-channel interference. But, because of the channel partitioning scheme and the symmetry in the system, we can reduce the co-channel interference by using the following channel reuse scheme.

In each cell, inner region uses the same band $W_1$. While $k^{th}$ FRN uses band $W_{2,k}$ to communicate with BS and MS in $k^{th}$ $A_2$ region uses $W_{3k}$ band to communicate with $k^{th}$ FRN.

### 2.3. Propagation Model

Wireless channel suffers from fading. Fading is mainly divided into two types, slow and fast. Slow fading is due to path-loss and shadowing, while fast fading is due to multi-path. In this paper, we assume that the code lengths are large enough to reveal the ergodic nature of fast fading. Hence, we do not explicitly consider multi-path effect. We focus on the path-loss and shadowing in the analysis. Because of the path-loss, the received signal power is inversely proportional to the distance between the transmitter and the receiver. In general, the path-loss $P_L$ between a transmitter and a receiver is given as,

$$P_L = \frac{P_T}{P_R} G_T G_R = \left(\frac{4\pi f}{c}\right)^2 \left(\frac{d^*}{d_0}\right)^\gamma \qquad (1)$$

where $P_T$ is the transmitted power; $P_R$ is the received power, $G_T$ and $G_R$ are the antenna gain of transmitter and receiver respectively; $f$ is the carrier frequency, $c$ is the speed of light; $d^*$ is separation between the transmitter and receiver; $d_0$ is the reference distance, and $\gamma > 0$ is the path-loss exponent [10].

## 3. RELAY PLACEMENT SCHEME

Improvement in capacity and increase in coverage area are the main benefits of FRNs. These benefits of FRNs are based on the position of relays in the cell. Deploying FRNs around the edge of the cell help the edge users. However, when they are placed at inappropriate locations, may cause interference to the edge users of the neighboring cell. Therefore, optimal placement of FRNs is a key design issue. Consider the downlink scenario where the BS encodes the message and transmits it in the first time slot to nearby MSs and FRNs. FRNs transmits the message to MSs at the cell boundary in the second time slot. FRNs are either Decode-and-Forward (DF) type, which fully decodes and re-encodes the message, or Amplify-and-Forward (AF) type, which amplifies and forwards the Message to MSs in the second hop. Note that the reverse will be for uplink scenario. In both uplink and downlink scenarios, we consider non-transparent type relays, i.e., MSs in the first hop communicate to BS while MSs in the second hop communicate only to FRNs.

### 3.1 Problem Formulation

In this section, we formulate the problem and describe our approach. Consider uplink scenario in cellular system as shown in Fig.1. In the cell, an MS can communicate directly with the BS through one-hop, or via FRN based on distance routing protocol [9]. Now, we assume that MSs are uniformly distributed. Hence, the average number of users in any region is proportional to its area. Let $N_1$ be the average number of MSs in $A_1$ (direct link) and $N_2$ be the average number of MSs in $A_2$ (relay link). Clearly, $N_1 = \lambda A_1$ and $N_2 = \lambda A_2$.

Let $R_1$, $R_2$ and $R_3$ be the data rate achieved on MS-BS (direct link), MS-FRN and FRN-BS links respectively. When MS transmits data to BS via FRN, the rate on MS-FRN link ($R_3$) should be equal to rate on FRN-BS link ($R_2$), i.e.,

$$R_2 = R_3,$$
$$W_2 log_2(1 + \Gamma_{RM}^a) = W_3 log_2(1 + \Gamma_{BR}^a) \qquad (2)$$

where $\Gamma_{RM}^a$ and $\Gamma_{BR}^a$ denote the worst case SIR for links MS to $k^{th}$ FRN and $k^{th}$ FRN to BS given that the MS is in any of the regions $A_{2k}$'s for k = 1, . . . , 6, respectively. Moreover, the data rate achieved on direct link (R1) is related to the rate on MS-FRN and FRN-BS ($R_2$ and $R_3$) as follows:

International Journal of Wireless & Mobile Networks (IJWMN) Vol. 4, No. 3, June 2012

$$\frac{R_1}{R_2} = \frac{N_1}{N_2} = \frac{A_1}{A_2} = \frac{A_1}{\sum_{i=1}^{6} A_{F,i}} = \frac{\left(\frac{d}{D}\right)^2}{1-\left(\frac{d}{D}\right)^2}, \qquad (3)$$

where $R_1 = W_1 log_2(1 + \Gamma_{BM}^a)$. $\Gamma_{BM}^a$ denotes the worst case SIR on a direct link MS to BS given that the MS is in the region $A_1$.

The worst case SIR is obtained by considering a scenario in which the transmitter of interest is placed at the longest possible distance from its receiver, while its interferers are placed at the shortest possible distance from the receiver. The worst case SIR of a link provides a lower bound on its actual SIR. Thus, to guarantee the required rate on a link, it suffices to guarantee the rate for the worst SIR on the link. We note that the resource allocation depending on the worst SIR may be conservative. But, it is conducive for robust and scalable implementation as the worst SIR does not change on account of system dynamics caused by the arrival and departure of users and also by the user mobility. The co-channel interference to the BS/FRN of interest is assumed to be from MSs or FRNs that links to first tier or upper tier cells (i[th] MS or k[th] FRN, the expression for the worst case SIRs of the three links is as follows.

$$\Gamma_{BM}^a = \frac{1}{d^{\gamma b}} \left[\sum_{i=1}^{N} \frac{1}{d_i^{\gamma b}}\right]^{-1} \qquad (4)$$

$$\Gamma_{BR}^a = \frac{1}{d_r^{\gamma r}} \left[\sum_{i=1}^{N} \frac{1}{d_{ri}^{\gamma r}}\right]^{-1} \qquad (5)$$

$$\Gamma_{RM}^a = \frac{1}{(d_m^{\gamma m})} \left[\sum_{i=1}^{N} \frac{1}{d_{mi}^{\gamma m}}\right]^{-1} \qquad (6)$$

where, d is the distance of furthest MS in $A_1$ from BS of interest ($BS_0$). Clearly, the distance between a MS in region $A_1$ and $BS_0$ is less than or equal to d while $d_i$ is the shortest possible distance between MS of i[th] interferer cell and $BS_0$. Similarly $d_r$ is the distance of k[th] FRN of $BS_0$ and $d_{ri}$ is the distance between k[th] FRN of i[th] interferer cell and the $BS_0$; $d_m$ is the furthest possible distance of MS from k[th] FRN within the coverage of $BS_0$ and $d_{mi}$ is the distance between MS that is within the coverage of k[th] FRN of i[th] interferer cell and k[th] FRN of $BS_0$. $\gamma b, \gamma r$ and $\gamma m$ are path-loss exponent of MS-BS, FRN-BS, and MS-FRN links, respectively.

The channel partitioning ($W_1$, $W_2$, and $W_3$) for the proposed scheme is obtained by solving Equ. (2) and (3). Let us define,

$$\Delta = \frac{\left(\frac{d}{D}\right)^2}{1-\left(\frac{d}{D}\right)^2} \frac{log_2(1+\Gamma_{RM}^a)}{log_2(1+\Gamma_{BM}^a)} + \frac{log_2(1+\Gamma_{RM}^a)}{log_2(1+\Gamma_{BR}^a)} + 1.$$

$$W_1 = \frac{log_2(1 + \Gamma_{RM}^a)}{log_2(1 + \Gamma_{BM}^a)} \frac{W}{\Delta}; \qquad (7)$$

$$W_2 = \frac{W}{\Delta}; \qquad (8)$$

$$W_3 = \frac{log_2(1 + \Gamma_{RM}^a)}{log_2(1 + \Gamma_{BR}^a)} \frac{W}{\Delta}. \qquad (9)$$

Equ.(7), (8) and (9) show that the channel partitioning in each FRN enhanced cell is dependent on the worst case SIRs of the three links which in turn depends on the FRN location (Equ.(4),

(5) and (6)). With this partitioning scheme, channel assignment to MSs can be conducted at both BS and FRNs. The same technique can be applied to share a channel among FRNs.

The aim of channel partitioning and relay positioning scheme is to maximize the number of users while satisfying the required rate of each user. Hence, maximizing the user density to support large number of users and determining the ratio of frequency band of three links are the key optimization problem. Therefore, the optimal user density of the inner and outer region based on SIR is formulated as,

$$\text{maximize } \lambda$$
$$s.t\ W_1, W_2 \text{ and } W_3 \text{ satisfying}$$
$$Equ. (7), (8) \text{ and } (9) \qquad (10)$$

Given that the relay nodes are placed at distance $d_r < D$, the worst case SIRs can be explicitly computed using Equ.(4), (5) and (6) and the geometry of the cellular system. Now, the goal is to find $d^*$ which maximizes $\lambda$. On solving the Equ. (10), determine $d^*$, which maximize the number of users supported in the system.

### 3.2 SIR Computation for the proposed scheme

We compute the SIR for each relay enhanced Cellular network. We consider the interference from the cells of two tires surrounding the $BS_0$ as shown in Fig.2. In this computation, the desired MS is assumed to be placed at the furthest distance from its BS/FRN while interferer MSs are located at the nearest possible locations to the $BS_0/FRN_1$ as shown in Fig.2. SIR for BS-MS (direct link) is given by

$$\Gamma_{BM}^a(d) = \frac{1}{d^{\gamma b}} \left[ \sum_{i=1}^{18} \frac{1}{d_i^{\gamma b}} \right]^{-1} = \frac{\left(\frac{d}{D}\right)^{-\gamma b}}{6\left(\left[\frac{\sqrt{3}}{2}\left(2-\frac{d}{D}\right)\right]^{-\gamma b} + \left[\frac{\sqrt{3}}{2}\left(4-\frac{d}{D}\right)\right]^{-\gamma b} + \left[\frac{\sqrt{3}}{2}\left(3-\frac{d}{D}\right)\right]^{-\gamma b}\right)} \qquad (11)$$

where $d_i$ is the distance between desired BS ($BS_0$) and interfering MSs of interfering BSs. Now, we consider the SIR computation of BS-FRN link where BS0 communicates with its $FRN_1$, the SIR for this link depends on the distance of $FRN_1$ of interferer cell from the $FRN_1$ of $BS_0$ as shown in Fig.4.

Let us define

$$\chi_{i,j} = (iD^2 + jd_rD + d_r^2 +)^{-\frac{\gamma r}{2}} + (iD^2 + jd_rD - d_r^2)^{-\frac{\gamma r}{2}}$$

$$\Gamma_{BR}^a(d_r) = \frac{1}{d_r^{\gamma r}} \left[ \sum_{i=1}^{N} \frac{1}{d_{ri}^{\gamma r}} \right]^{-1} = d_r^{-\gamma r} \left\{ (3D + d_r)^{\gamma r} + (3D - d_r)^{\gamma r} + 2(3D^2 + d_r^2)^{-\frac{\gamma r}{2}} + 2\left[\chi_{3,3} + \chi_{12,6} + \chi_{9,3} + (12D^2 + d_r^2)^{-\frac{\gamma r}{2}}\right] \right\}^{-1} \qquad (12)$$

Now, we compute the SIR on FRN-MS link. When the interference of the FRN is of concern, the interference is contributed by MSs located in other cells that are using the same channel to communicate with their corresponding FRNs. Fig.4 shows a case where $FRN_1$ of $BS_0$ communicate with its MS so that the signal from MS of other $FRN_1$ will be the source of the interference. SIR for this particular scheme is,

$$\Gamma_{RM}^a(d_m) = \frac{1}{d_m^{\gamma m}} \left[ \sum_{i=1}^{18} \frac{1}{d_{mi}^{\gamma m}} \right]^{-1} = \frac{\left(\frac{d_m}{D}\right)^{-\gamma m}}{6\left(\left[\frac{\sqrt{3}}{2}\left(1+\frac{d}{D}\right)\right]^{-\gamma m} + \left[\frac{\sqrt{3}}{2}\left(3+\frac{d}{D}\right)\right]^{-\gamma m} + \left[\frac{\sqrt{3}}{2}\left(2+\frac{d}{D}\right)\right]^{-\gamma m}\right)} \qquad (13)$$

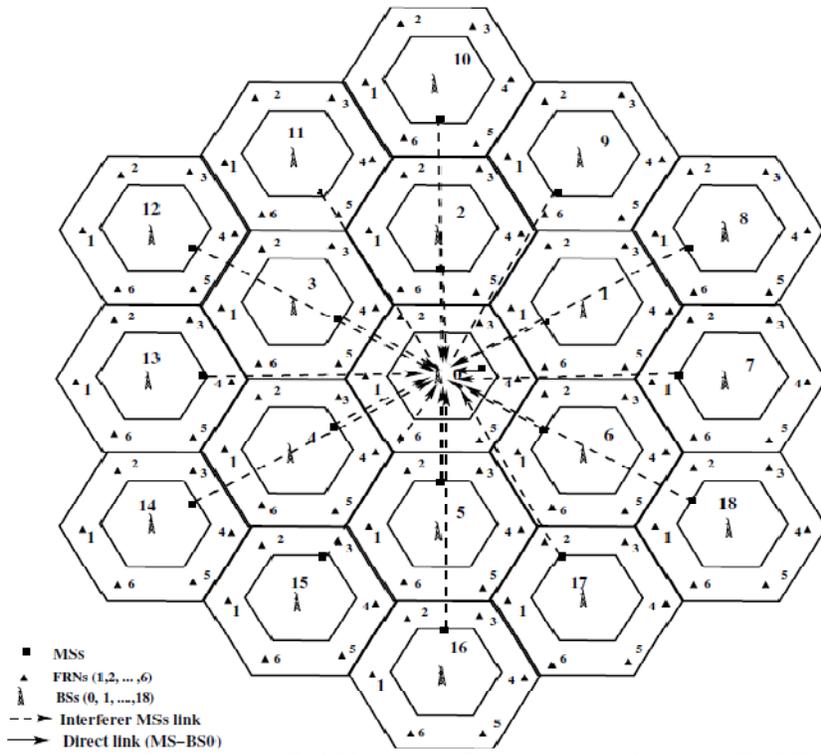

Fig.2. Worst-case interference on MS-BS link (interference received from MS of 18 co-channel cells)

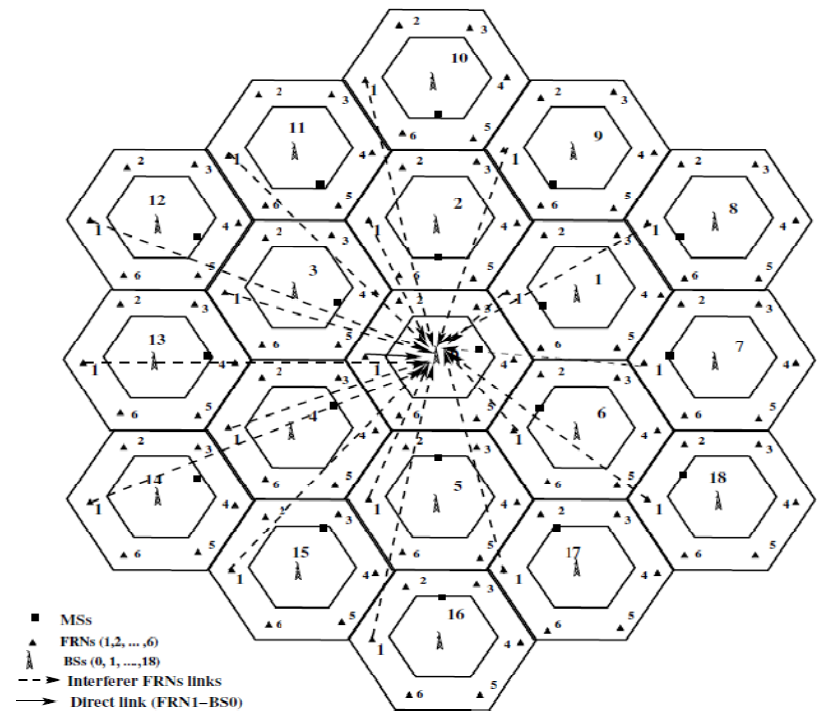

Fig.3. Worst-case interference on BS-FRN link (interference received from FRN1 of 18 co-channel cells)

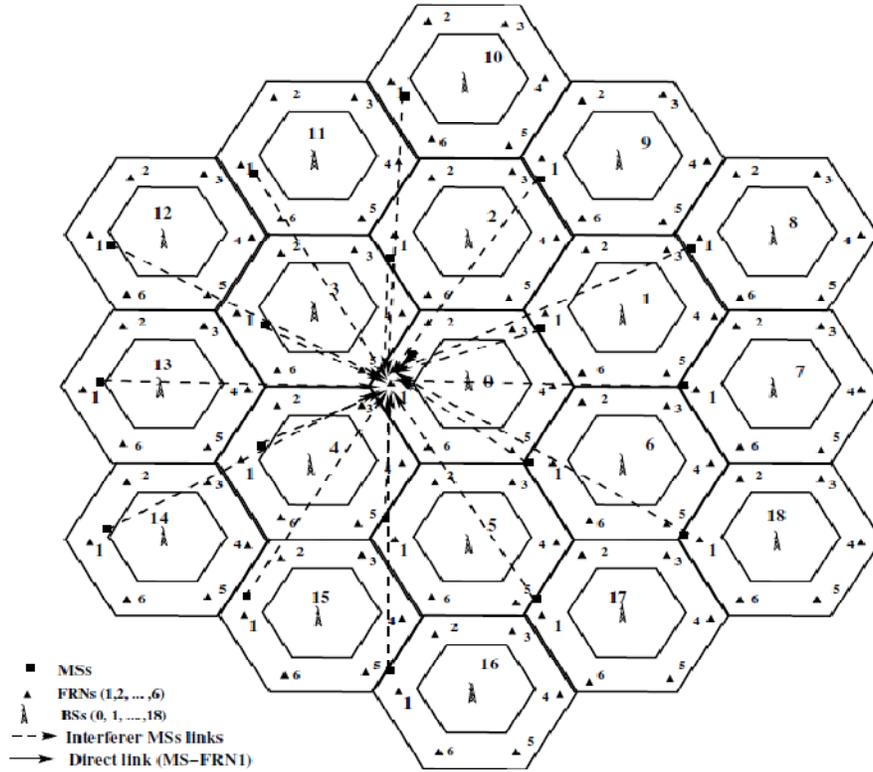

Fig.4 Worst-case interference on MS-FRN link (interference received from MSs in the coverage are of FRN1 of 18 co-channel cells).

### 3.3. Outage performance analysis

Co-channel interference and shadowing effects are among the major factors that limit the capacity and link quality of a wireless communications system. In this section, we use Gauss-Markov model to evaluate the statistical characteristics of SIR in Multi-hop communication channels. By modeling SIR as log-normally distributed random variable, we investigate the performance of relay placement scheme discussed in Section 3.1. We make a comparison between the above models in terms of performance evaluation where outage probability is a QoS parameter. Further more the result is used to find the more realistic way of channel partitioning and relay FRN placement schemes.

The co-channel uplink interference to the BS/FRN of interest is assumed to be from MSs or FRNs that links to first tier or upper tier cells ($MS_i$'s or $FRN_K$'s). Including the shadowing effect on the three links, the SIR on each link can be described as,

$$\Gamma_{BM}^b = \frac{10^{\xi b/10}}{d^{\gamma b}} \left[\sum_{i=1}^{N} \frac{1}{d_i^{\gamma b} 10^{\xi bi/10}}\right]^{-1} = \left[\sum_{i=1}^{N} \left(\frac{d}{d_i}\right)^{\gamma b} 10^{\frac{\xi bi - \xi b}{10}}\right]^{-1} \quad (14)$$

$$\Gamma_{BR}^b = \frac{10^{\xi r/10}}{d_r^{\gamma r}} \left[\sum_{i=1}^{N} \frac{1}{d_{ri}^{\gamma r} 10^{\xi ri/10}}\right]^{-1} = \left[\sum_{i=1}^{N} \left(\frac{d}{d_{ri}}\right)^{\gamma r} 10^{\frac{\xi ri - \xi r}{10}}\right]^{-1} \quad (15)$$

$$\Gamma_{RM}^b = \frac{10^{\xi m/10}}{d_m^{\gamma m}} \left[\sum_{i=1}^{N} \frac{1}{d_{mi}^{\gamma r} 10^{\xi mi/10}}\right]^{-1} = \left[\sum_{i=1}^{N} \left(\frac{d}{d_{mi}}\right)^{\gamma m} 10^{\frac{\xi mi - \xi m}{10}}\right]^{-1} \quad (16)$$

where d and $d_r$ are the location of desired MS and FRN from the $BS_0$ on direct link, $d_m$ is the location of desired MS from desired FRN under the second hop. Similarly, $d_i$ and $d_{ri}$ are the location of co-channel interferer MS and FRN from the $BS_0$, while $d_{mi}$ denotes the location of co-channel interferer MSs from FRN that is associated to $BS_0$. Shadowing for the desired links are denoted as $\xi d$, $\xi r$ and $\xi m$. For interfering links shadowing is expressed as $\xi di$, $\xi ri$ and $\xi mi$ to denote the interfering link of MS-BS, FRN-BS and MS-FRN; in all cases, $i \epsilon \{1,...,18\}$. Let the threshold SIR on BS-MS, BS-FRN and FRN-MS links are denoted as $\Gamma_b$, $\Gamma_r$ and $\Gamma_m$ respectively. The outage probability on these three links is given as,

$$P_b^{out} = \Pr(\Gamma_{BM}^b < \Gamma_b) = \int_0^{\Gamma_b} \frac{1}{\sigma_B\sqrt{2\pi}} exp\left[\frac{-(x-m_{\Gamma_B})^2}{2\sigma_B^2}\right] dx = 1 - Q\left(\frac{\Gamma_b - m_{\Gamma_B}}{\sigma_B}\right) \quad (17)$$

where Q(.) is the Gaussian function. We compute the mean $m_{\Gamma_B}$ and standard deviation $\sigma_B$ of SIR based on Fenton–Wilkinson's and Schwartz–Yeh's method [12]. Similarly, the outage probability on FRN-BS and MS-FRN are computed as $P_r^{out} = \Pr(\Gamma_{BR}^b < \Gamma_r)$ and $P_m^{out} = \Pr(\Gamma_{RM}^b < \Gamma_b)$ respectively. The equivalent SIR experienced by users over two-hop transmission depends on the type of relaying scheme. In this paper we consider Decode-and-Forward Relaying (DF) and Amplify-and-Forward Relaying (AF) relaying schemes.

### 3.3.1. Decode and Forward Relaying

Consider an uplink scenario, where each FRN decodes the signal received from MS-FRN link and re-transmits to BS on the FRN-BS link. In this scheme, the end to end rate achieved from BS to MS is determined by the minimum rate achieved among the rates of MS-FRN-BS and links, i.e., if the required rate for this scheme is $R_{DF}$, then

$$2R_{DF} \leq \min\{log_2(1+\Gamma_{BR}^b), log_2(1+\Gamma_{RM}^b)\} \quad (18)$$

If the required rate is not achieved on either of the link, then user is said to be in outage. Let the outage probability in DF scheme be $P_{DF}^{out}$, it can be expressed as

$$P_{DF}^{out} = \Pr(\min\{log_2(1+\Gamma_{BR}^b), log_2(1+\Gamma_{RM}^b)\} \leq 2R) \quad (19)$$

Equivalently, the outage probability on MS-FRN and FRN-BS links can be computed as

$$P_{DF}^{out} = \Pr(\Gamma_{BR}^b < \Gamma_r)\Pr(\Gamma_{RM}^b < \Gamma_m) = P_r^{out}.P_m^{out} \quad (20)$$

### 3.3.2. Amplify and Forward Relaying

In amplify and forward relaying scheme, the AF relay amplifies the analog signal received from the BS and transmits an amplified version of it to the MSs [14]. Let the experienced SIR for this scheme be $\Gamma^{AF}$, then it is computed as follows.

$$\Gamma^{AF} = \frac{\Gamma_{BR}^b \Gamma_{RM}^b}{\Gamma_{BR}^b + \Gamma_{RM}^b + 1} \quad (21)$$

Let $\Gamma$ be the threshold SIR. If the experienced SIR falls below the threshold, the user is said to be in outage and outage probability is given as $P_{AF}^{out}$ out for AF scheme [14].

$$P_{AF}^{out} = Pr\left(\frac{\Gamma_{BR}^b \Gamma_{RM}^b}{\Gamma_{BR}^b + \Gamma_{RM}^b + 1} \leq \Gamma\right) \quad (22)$$

Note that Shadowing is usually represented by i.i.d. log-normal model in wireless multi-hop models. However, shadowing paths are correlated. In this scheme, we consider correlation among interferers and also the correlation that may exist between interferer and desired signals.

### 3.3.3. Sectoring the inner region of the cell

Sectoring the inner region of FRN enhanced cell can efficiently reduce the co-channel interference. In sectoring, the total subcarriers in the inner region will be further divided among the number of sectors in the inner region. For example, if $60^0$ sectoring is applied, $W_1$ is partitioned into six orthogonal segments, viz. $W_{1,i}$ for i = 1, . . . , 6. MSs present in any sector of the inner region can use dedicated subcarriers allocated to that sector. Hence, these orthogonal sub-channels greatly reduce the inter-cell interference. As it was explained in Section 3.1, the total band allocated to inner region depends on the FRN placement scheme.

## 4. RESULTS AND DISCUSSION

In this section, we present both analytical and simulation results to illustrate the performance of our proposed algorithms. We use Matlab for modeling a cellular network under varying channel conditions. We analyze the performance of channel partitioning and relay placement schemes described in previous sections. The parameters used for simulations are shown in Table 1.

Table 1. List of the simulation parameters

| Parameters | Values |
| --- | --- |
| Carrier Frequency | 5GHz |
| System Bandwidth (W) | 25.6MHz |
| Cell Radius (D) | 1000m |
| Standard Deviation ($\sigma d$, $\sigma r$ and $\sigma m$) | 8, 5, 8 |
| Correlation Coefficient | 0.5 |
| Path-loss Exponent ($\gamma d$, $\gamma r$ and $\gamma m$) | 3.5, 2.5, 3.5 |
| BS Transmit Power ($P_{BS}$) | 40dBm |
| FRN Transmit Power ($P_{FRN}$) | 20dBm |
| MS transmit Power ($P_{MS}$) | 2dBm |
| Threshold ($\Gamma$) | -10dBm |
| Thermal Noise (N) | -100dBm |

Fig.5 shows the uplink SIR experienced by MS as it moves from the center to the boundary of the cell. The SIR experienced by the MSs of inner region (MS-BS) decreases monotonically while the SIR experienced by the MSs of relaying region (MS-FRN) increases such that the decrement in SIR of the inner region is compensated by SIR experienced by the MSs of outer region. The SIR experienced on the BS-FRN link is fairly constant over the entire range of d/D. This figure, demonstrates that FRN enhanced cellular system offers remarkable improvement in SIR experienced by users over the conventional system without relays. Fig.6 shows the variation of total data rate of the inner (R1) and outer region (R2) as a function of d/D for the path-loss exponent of 3. From this figure, we observe that FRN positioning has significant impact on the performance of FRN based OFDMA cellular network. In addition, the data rate is also dependent on environmental conditions expressed by path-loss exponent.

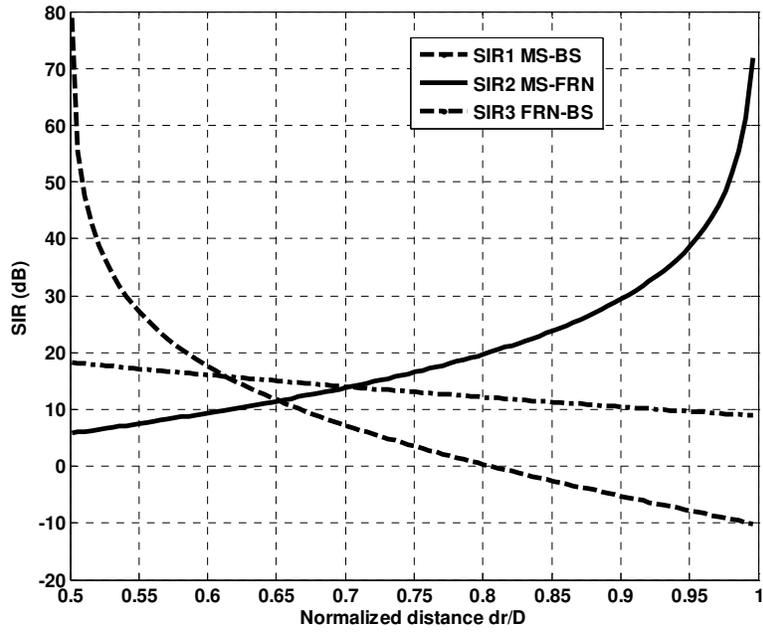

Fig.5 Distribution of SIRs versus dr/D

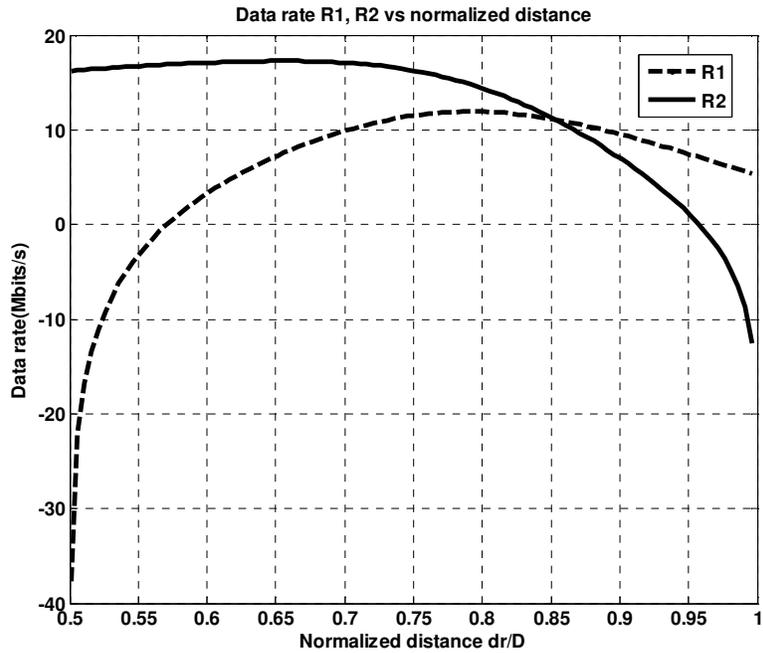

Fig.6 Cell capacity versus dr/D

31

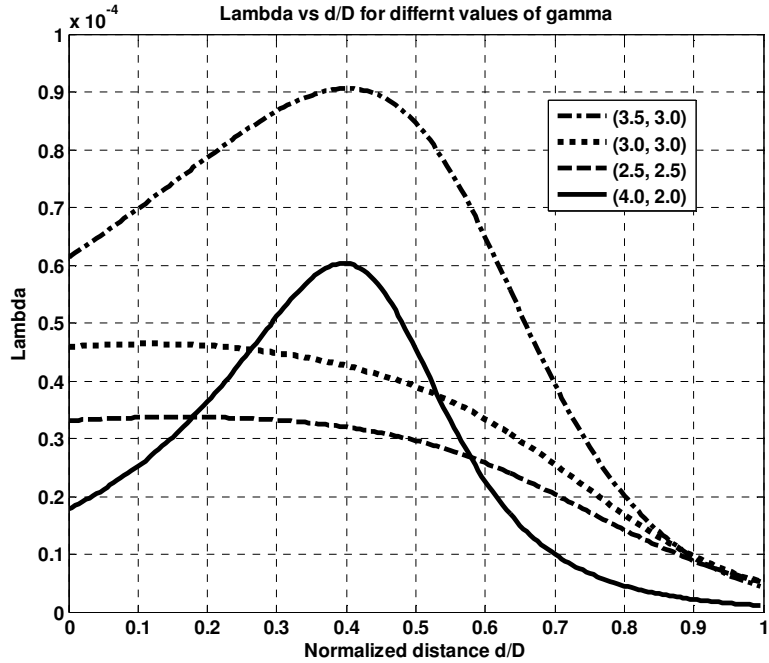

Fig.7 User density for pairs of path-loss exponent (for MS-BS/FRN and FRN-BS link)

Fig.7 shows locations of maximum user density for a given path-loss exponent that indicates the optimal FRN position under different path-loss conditions. The plots are denoted by pair of path-loss exponent of two links. The first number indicates the path-loss exponent of MS-BS/FRN link and second indicates that of FRN-BS link. From this figure we observe that for path-loss exponent of (3.5, 3) and (4, 2), the optimal relay node location is at 70% of the cell radius. In case of (3, 3) and (2.5, 2.5), the optimal FRN position is 55% of the cell radius. It can be seen from this figure that the optimal FRN location depend on path-loss exponents which contradict with the assumption of FRN location as dr/D = 2/3, i.e., d/D = 1/3 in [8] and other literatures.

Fig.8 shows the partitioning of available uplink frequency band ($W_1$, W2 and $W_3$) in each FRN enhanced cell to serve three set of links (MS-BS, FRN-BS and MS-FRN links). From this graph, we observe that the amount of band shared by these three set of links is dependent on FRN location. Such spectrum partitioning scheme can be conducted at both BS and FRNs. Fig.9 shows the effect of path-loss exponent on channel band partitioning for a fixed distance of MS and FRN/BS. We consider path-loss exponents of 2.0 and 2.5 for FRN-BS link and determine the partition ($W_1$, $W_2$, and $W_3$). For a fixed radial distance of a user from BS, the SIR experienced by user decreases as path-loss exponent increases. This intern affect the frequency band allocation to each link as shown in Fig.9.

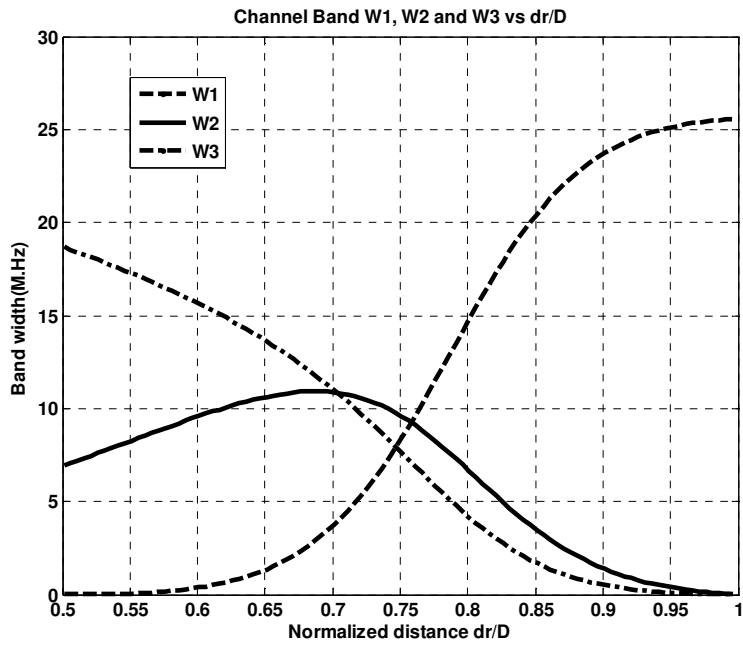

Fig.8 Effect of dr/D on channel Band

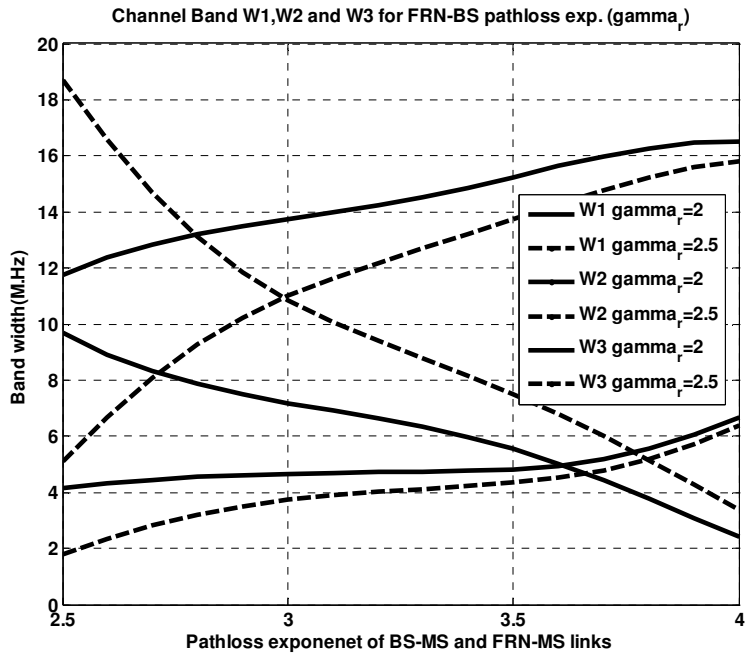

Fig. 9 Effect of path-loss exponent on channel Band

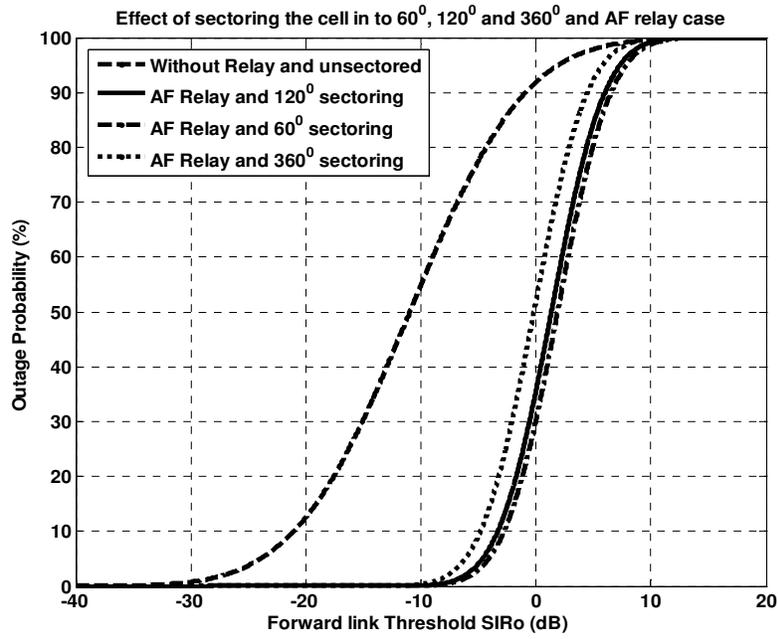

Fig.10. Outage probability AF Relay compared with cellular networks without relay; the inner region of relay enhance cell with $60^0$, $120^0$ and $0^0$ sectoring.

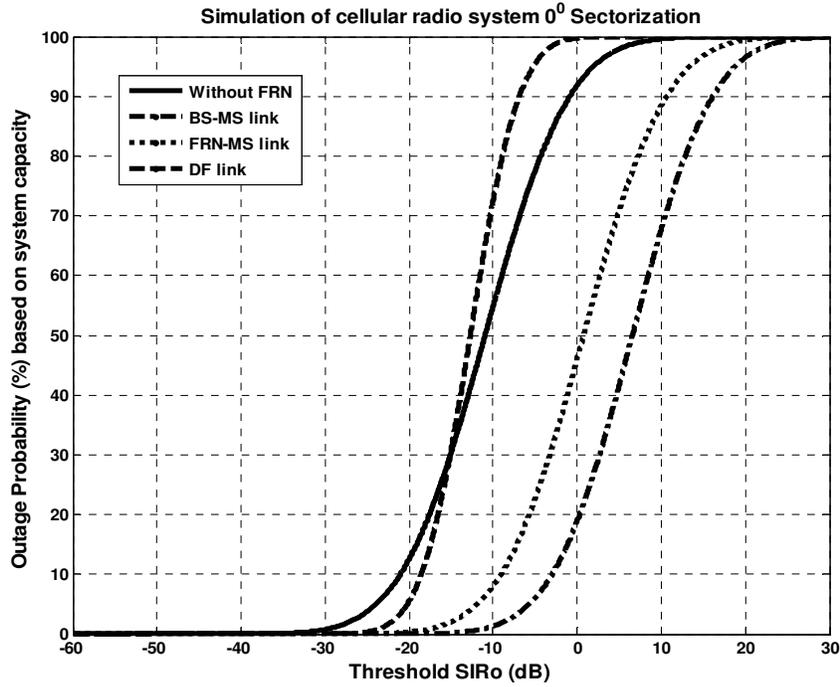

Fig.11. Outage probability of DF Relay compared with cellular networks without relay; no sector in inner region.

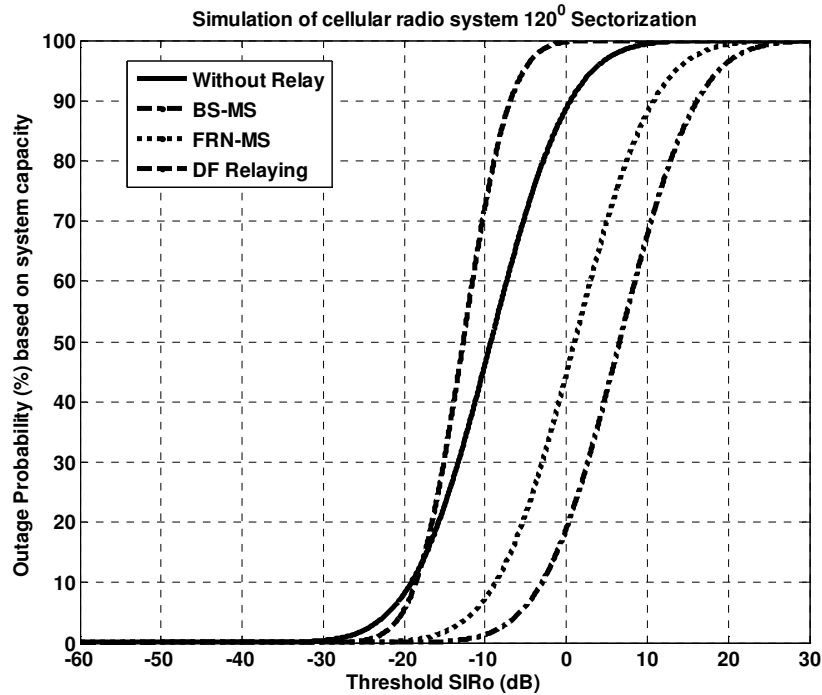

Fig.12. Outage probability of DF Relay compared with cellular networks without relay; $120^0$ sector in the inner region.

Fig.10 shows the outage probability versus SIR in dB for an AF relay and without relay. From the figure, we observe that AF relay scheme achieves significant improvement on the performance of the cellular network. It reduces the outage probability significantly. For example, at the threshold SIR of 0 dB, the AF scheme can improve the outage probability from 90% to 30%. In addition, sectoring of the inner cell also contribute in the overall system improvement

Fig. 11 shows the outage probability for DF relays. For lower value of threshold SIR, both AR and DF relays show significant improvement on the outage performance of the cell. However, for the higher values of SIR threshold limit, the outage probability increases. This shows that the system is dominated by interference. At lower region of SIR threshold, AF relaying scheme performs better than DF in improving the outage probability of the users in terms of link quality. Fig.11 and Fig.12 compares the outage probability of a cell using DF relaying scheme by sectoring the inner region of the cell. Even though sectoring improves the outage probability of users, there is a trade-off that it may degrade the capacity of a system in terms of maximum carried traffic.

## 5. CONCLUSION

In this work we propose new techniques for channel partitioning and FRN positioning schemes in cellular OFDMA network. We investigate channel partitioning and FRN placement under different path loss conditions. Our analytical results show that the optimal relay node placement depends on the path-loss exponent of the environment. For example it is at 0.7 of dr/D ratio for path-loss exponent of (3, 3). In addition to this, simulation results show that our proposed schemes improve the system capacity and provides better QoS for users present at the boundary of cells.

## ACKNOWLEDGMENT

This research work is supported by India-UK Advanced Technology Centre (IU-ATC) of Excellence in Next Generation Networks Systems and Services.

**References**

[1]  Zaher Dawy, Sami Arayssi, Ibrahim Abdel Nabi, and Ahmad Husseini. Fixed relaying with advanced antennas for CDMA Cellular Networks. In IEEE GLOBECOM proceedings, 2006.

[2]  H. Hu, H. Yanikomeroglu, and et al. Range extension without capacity penalty in Cellular Networks with digital fixed relays. IEEE Globecom, Dec 2004.

[3]  I. Katzela and M. Naghshineh. Channel assignment scheme for cellular mobile telecommunication systems: A comprehensive survey. IEEE Personal Communication, pages 10–31, June 1996.

[4]  Jian Liang, Hui Yin, Haokai Chen, Zhongnian Li, and Shouyin Liu. A novel dynamic full frequency reuse scheme in OFDMA cellular relay networks. In Vehicular Technology Conference (VTC Fall), 2011 IEEE, pages 1 –5, Sept. 2011.

[5]  Min Liang, Fang Liu, Zhe Chen, Ya Feng Wang, and Da Cheng Yang. A novel frequency reuse scheme for OFDMA based Relay enhanced cellular networks. In Vehicular Technology Conference, 2009. VTC Spring 2009. IEEE 69th, pages 1 –5, April 2009.

[6]  Sultan F. Meko and Prasanna Chaporkar. Channel Partitioning and Relay Placement in Multi-hop Cellular Networks. In Proc. IEEE ISWCS, 7-10 Sept. 2009.

[7]  S. Mengesha, H. Karl, and A. Wolisz. Capacity increase of multi-hop cellular WLANs exploiting data rate adaptation and frequency recycling. In MedHocNet, June 2004.

[8]  M. Rong et al P. Li. Reuse partitioning based frequency planning for relay enhanced cellular system with NLOS BS-Relay links. IEEE, 2006.

[9]  V. Sreng, H. Yanikomeroglu, and D. D. Falconer. Relayer selection strategies in cellular networks with peer-to-peer relaying. IEEE, VTC, 03, Oct 2003.

[10] D. Tse and P. Viswanath. Fundamentals of Wireless Communications. Cambridge University Press, 2005.

[11] D. Walsh. Two-hop relaying in CDMA networks using unlicensened bands. Master's thesis, Carleton Univ., Jan 2004.

[12] Jingxian Wu, N.B. Mehta, and Jin Zhang. Flexible lognormal sum approximation method. In EEE GLOBECOM, pages 3413–3417, Dec 2005.

[13] H. Yanikomeroglu. Fixed and mobile relaying technologies for cellular networks. In Second Workshop on Applications and Services in Wireless Networks, July 2002.

[14] Bhuvan Modi1, A. Annamalai1, O. Olabiyi1 and R. Chembil Palat. Ergodic capacity analysis of cooperative Amplify-and-Forward relay networks over rice and nakagami fading channels. International Journal of Wireless & Mobile Networks (IJWMN) Vol. 4, No. 1, February 2012.

**Author Biography**

He is Research scholar in Electrical Engineering at Indian Institute of Technology (IIT) Bombay, India. His research interests include Resource allocation, Scheduling and Queuing model in wireless networks.

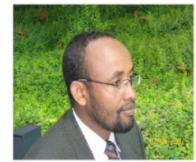